\documentclass[12pt]{article}

\usepackage{color}

\usepackage{graphicx}
\usepackage{amssymb,amsmath,array}

\newcommand{\tz}{{\tilde z}}
\newcommand{\tw}{{\tilde w}}
\newcommand{\pz}{\partial_{z}}
\newcommand{\ptz}{\partial_{\tilde z}}
\newcommand{\pw}{\partial_{w}}
\newcommand{\ptw}{\partial_{\tilde w}}
\newcommand{\Az}{A_{z}}
\newcommand{\Atz}{A_{\tilde z}}
\newcommand{\Aw}{A_{w}}
\newcommand{\Atw}{A_{\tilde w}}

\begin{document}

\begin{center}
{\bf\Large Bianchi permutability for the\\ \vskip 1mm anti-self-dual Yang-Mills equations}

\vskip 7 mm

{\large G. Benincasa and R. G. Halburd}

\vskip 4 mm

Department of Mathematics,
University College London
\\
Gower Street,
London WC1E 6BT
\end{center}

\vskip 6 mm

\begin{abstract}
The anti-self-dual Yang-Mills equations are known to have reductions to many integrable differential equations.  
A general B\"acklund transformation (BT) for the ASDYM equations generated by a Darboux matrix with an affine dependence on the spectral parameter is obtained, together with its Bianchi permutability equation.
We give examples in which we obtain BTs of symmetry reductions of the ASDYM equations by reducing this ASDYM BT.
Some discrete integrable systems are obtained directly from reductions of the ASDYM Bianchi system.
\end{abstract}

\vskip 5 mm

\noindent\centerline{\em Dedicated to Mark Ablowitz on his 70th birthday}

\section{Introduction}
It is well known that many discrete integrable equations arise from the permutability of B\"acklund transformations of continuous integrable systems.
In turn, many continuous integrable systems are known to be symmetry reductions of the anti-self-dual Yang-Mills (ASDYM) equations
\cite{ward:77,chakravartyac:90,ablowitzc:91,masonw:96,ablowitzch:03}.
In this paper we will derive a form of the B\"acklund transformations for the ASDYM equations that is well-suited for obtaining B\"acklund transformations of reductions.  Furthermore, we will use the Bianchi permutability of these B\"acklund transformations to obtain discrete integrable equations.

Let  ${A}:=\Az\,{\rm d}z+\Aw\,{\rm d}w+\Atz\,{\rm d}\tz+\Atw\,{\rm d}\tw$ be a one-form with components in some Lie algebra $\mathfrak{g}$.  In all of our examples, $\mathfrak{g}$ will be $\mathfrak{sl}(2;{\mathbb C})$.
In terms of the components of this one-form, the ASDYM equations with respect to the metric ${\rm d}s^2= 2({\rm d}z\,{\rm d}\tilde z-{\rm d}w\,{\rm d}\tilde w)$ are
\begin{equation}
\label{asdym}
\begin{split}
&
{\pz\Aw-\pw\Az+[\Az,\Aw]=0},
\\
&
{\ptz\Atw-\ptw\Atz+[\Atz,\Atw]=0},
\\
&
{\pz\Atz-\ptz\Az-\pw\Atw+\ptw\Aw+[\Az,\Atz]-[\Aw,\Atw]=0}.
\end{split}
\end{equation}
This is a system of three equations in four $\mathfrak{g}$-valued functions $\Az$, $\Aw$, $\Atz$ and $\Atw$ of the four variables $z$, $\tilde z$, $w$ and $\tilde w$.  We will allow these independent variables to be complex.
The system (\ref{asdym}) is the compatibility condition for the Lax pair
\begin{equation}
\label{lax}
\begin{split}
{(\pz-\zeta\ptw)\Psi }&=-(\Az-\zeta\Atw)\Psi,\\
(\pw-\zeta\ptz)\Psi &=-(\Aw-\zeta\Atz)\Psi.
\end{split}
\end{equation}
This Lax pair is equivalent to the statement that the differential operators
$L=D_w-\zeta D_\tz$
and
$M=D_z-\zeta D_\tw$
commute, where
$D_z=\pz+\Az$,
$D_w=\pw+\Aw$,
$D_\tz=\ptz+\Atz$, and
$D_\tw=\ptw+\Atw$.

The conformal symmetries of ${\mathbb R}^{2,2}$ (translations, rotations/boosts, dilations, inversions) induce (Lie point) symmetries of the ASDYM equations.  This large group of symmetries gives rise to a rich collection of inequivalent reductions.
The ASDYM equations are also invariant under the gauge transformation
$$
A_\mu\mapsto g^{-1}\partial_\mu g+g^{-1}A_\mu g,
$$
for any G-valued function $g$, where $G$ is the Lie group of $\mathfrak{g}$.  The gauge freedom can be used to remove arbitrary functions in reductions.

The ASDYM equations also admit a variety of B\"acklund transformations, which in general are non-point symmetries in that they depend on derivatives of solutions and not just values of the solutions.
B\"acklund transformations for the (A)SDYM equations have been introduced by several authors including
Corrigan, Fairlie, Yates and Goddard \cite{corriganfyg:78},
Prasad, Sinha and Chau Wang \cite{prasadsw:79},
Bruschi, Levi and Ragnisco \cite{bruschilr:82},
Papachristou and Kent Harrison \cite{papachristoukh:87},
Mason, Chakravarty and Newman \cite{masoncn:88}, and 
Tafel \cite{tafel:90}.

Despite the fact that many integrable systems are known to be reductions of the ASDYM equations, there are very few instances where BTs for the reduced equations have been obtained as reductions of an appropriate BT for ASDYM equations.
Masuda obtained the affine Weyl group symmetry of $P_{\rm II}$,  $P_{\rm III}$ and  $P_{\rm IV}$ from the ASDYM B\"acklund transformation \cite{masuda:05,masuda:07}.

One of the technical difficulties in reducing the BT for the ASDYM equations is the need to change gauge.
In this paper we will derive the general form of a B\"acklund transformation generated by a Darboux matrix with an affine dependence on the spectral parameter.  This form of the B\"acklund transformation makes it much easier to deal with issues related to gauge.

The first two of the 
equations in (\ref{asdym})
guarantee the existence of $G$-valued functions $H$ and $K$ respectively such that
$\pz H=- A_zH$,
$\pw H=- A_wH$,
$\ptz K=- \Atz K$,
and $\ptw K=- \Atw K$.
The final equation,
$${
\pz\Atz-\ptz\Az-\pw\Atw+\ptw\Aw+[\Az,\Atz]-[\Aw,\Atw]=0},
$$
then takes the compact form
\begin{equation}
\label{yang}
\pw(J^{-1}\ptw J)-\pz (J^{-1}\ptz J)=0,
\end{equation}
where
$J=K^{-1}H$.  This is known as {\em Yang's equation} and was first written in \cite{yang:77}.
Equation (\ref{yang}) has the obvious symmetry
$$
J(z,w,\tilde z,\tilde w)\mapsto M(z,w) J(z,w,\tilde z,\tilde w) \widetilde M(\tz,\tw).
$$

Apart from reductions to the Ernst equation and the chiral fields models, this form of ASDYM is not usually considered in relation to symmetry reductions.
One reason for this is that Yang's equation has lost some of the Lie symmetries of the original system (the rotations and boosts).
Another reason is that the $A_\mu$, in the original formalism appear directly in the Lax pair of ASDYM, which makes it easier to identify reductions by looking at the reduced Lax pair.

In section 2 we will derive a B\"acklund transformation for the ASDYMEs.  This system is most naturally expressed in terms of Yang's equation (\ref{yang}).  The equations describing these transformations depend on two matrix-valued functions, $C(z,w)$ and $\widetilde C(\tilde z,\tilde w)$.  These matrices contain important gauge information.  This form of the BT is particularly useful when considering reductions.  The case in which $C$ and $\widetilde C$ are constants was derived previously by Bruschi, Levi and Ragnisco  \cite{bruschilr:82}.  In section 3 we consider the known reductions of the ASDYMEs to the sine-Gordon equation and the sixth Painlev\'e equation.  We show how these reductions can be extended to the BTs derived in section 2 to obtain the B\"acklund transformations for the reductions.  Finally, in section 4, we derive an equation describing the Bianchi permutability of the BTs for the ASDYMEs derived in section 2.  We show how reductions of this system lead to discrete integrable equations.

\section{B\"acklund transformations  for ASDYM equations}

Starting from the ASDYM Lax pair (\ref{lax}),
we perform a $\zeta$-dependent gauge transformation
\begin{equation}
\label{darboux}
\Psi\mapsto \hat\Psi=(S+\zeta T)\Psi,
\end{equation}
such that the resulting system has the same form:
\begin{equation*}
\begin{split}
{(\pz-\zeta\ptw)\hat\Psi }&{=-(\hat\Az-\zeta\hat A_\tw)\hat\Psi},\\
{(\pw-\zeta\ptz)\hat\Psi }&{=-(\hat\Aw-\zeta\hat A_\tz)\hat\Psi}.
\end{split}
\end{equation*}
This gives
\begin{equation*}
\begin{split}
\big\{(S_w-S\Aw+{\hat A}_w S)+\zeta(T_w-S_{\tilde z}+S\Atz-T\Aw+{\hat A}_wT-{\hat A}_{\tilde z}S) &\\
+\zeta^2(-T_{\tilde z}+TA_{\tilde z}-\hat A_{\tilde z}T)\big\}\Psi &=0,\\
\big\{(S_z-S\Az+\hat\Az S)+\zeta(T_z-S_{\tilde w}+S\Atw-T\Az+{\hat A}_z T-{\hat A}_{\tilde w}S) &\\
+\zeta^2(-T_{\tilde w}+TA_{\tilde w}-\hat A_{\tilde w}T)\big\}\Psi &=0.
\end{split}
\end{equation*}
Equating the coefficients of the various powers of $\zeta$ yields
\begin{equation}
\label{STeqns}
\begin{split}
S_w &=S\Aw-{\hat A}_w S,
\qquad
S_z = S\Az-\hat\Az S,\\
S_{\tilde z}-T_w &= S\Atz-{\hat A}_{\tilde z}S-T\Aw+{\hat A}_wT,
\\
S_{\tilde w}-T_z&=S\Atw-{\hat A}_{\tilde w}S-T\Az+{\hat A}_z T,
\\
T_{\tilde z} &=TA_{\tilde z} -\hat A_{\tilde z}T,
\qquad
T_{\tilde w} = TA_{\tilde w}-\hat A_{\tilde w}T.
\end{split}
\end{equation}
Recall that two of the three ASDYM equations guarantee the existence of the $G$-valued potential functions $H$ and $K$ such that
$\pz H=- A_zH$,
$\pw H=- A_wH$,
$\ptz K=- \Atz K$,
and $\ptw K=- \Atw K$.  Similarly there are functions $\hat H$ and $\hat K$ such that $\pz \hat H=- \hat A_z\hat H$, $\ptz \hat K=- \hat {A}_{{\tilde z}} \hat K$, etc.
Now define
${C:={\hat{K}}^{-1}TK}$ and
${\tilde C:={\hat{ H}}^{-1}S H}$.
Then the first two equations in (\ref{STeqns}) are equivalent to
${\widetilde C\equiv \widetilde  C(\tilde z,\tilde w)}$ and the last two equations  in (\ref{STeqns}) are equivalent to ${C\equiv C(z,w)}$.
In terms of $J=K^{-1}H$, the remaining two equations in (\ref{STeqns}) become
\begin{equation}
\label{BT}
\begin{split}
\hat J\left( \hat J^{-1}C J\right)_z &= \left( \hat J\widetilde C J^{-1}\right)_{\!\tilde w}J,
\\
\hat J\left( \hat J^{-1}C J\right)_w &= \left( \hat J\widetilde C J^{-1}\right)_{\!\tilde z}J.
\end{split}
\end{equation}
In the case when $C$ and $\widetilde C$ are constant multiples of the identity, this system was first derived by Prasad, Sinha and Chau Wang \cite{prasadsw:79}.
The case in which in which $C$ and $\widetilde C$ constant was later derived by
Bruschi, Levi and Ragnisco  \cite{bruschilr:82}.   The more general dependence 
${C\equiv C(z,w)}$ and ${\widetilde C\equiv \widetilde  C(\tilde z,\tilde w)}$ is needed to ensure that $\hat J$ remains in $G$ (e.g., if $G={\rm SL}(2;{\mathbb C})$ then we need this freedom $C$ and $\widetilde C$ to ensure that ${\rm det}(\hat J)=1$) and plays an important role in some reductions.

\section{Examples of reductions}
We review some standard reductions of the ASDYM equations in the form (\ref{asdym}) and extend these reductions to include B\"acklund transformations.

\subsection{Reduction to  the sine-Gordon equation}

Following Chakravarty and Ablowitz \cite{chakravartya:92}, we look for solutions of the ASDYM equations such that the $A_\mu$'s depend on $z$ and $\tz$ only.  
We choose a gauge such that $\Atz=0$.
The field equations (\ref{asdym}) become
$$
\pz\Aw+[\Az,\Aw]=0,\quad
\ptz \Atw=0,\quad\mbox{and}\quad
\ptz\Az+[\Aw,\Atw]=0.
$$
Generically, we can use the remaining gauge freedom to put $\Atw$ in the form
$$
k\left(
	\begin{matrix}
		0 & 1\\
		1 & 0
	\end{matrix}
\right).
$$
For the remaining matrices, we take 
$$
\Aw=
\left(
	\begin{matrix}
		0 & a-ib\\
		a+ib & 0
	\end{matrix}
\right)
\quad\mbox{and}\quad
\Az=
\left(
	\begin{matrix}
		c & 0\\
		0 & -c
	\end{matrix}
\right).
$$
This leads to the equations

\centerline{
$
a_z=2ibc$,
$
b_z=-2iac$
\ \ and\ \ 
$
c_{\tz}=2ikb.
$
}

\noindent
The first two equations give
$a^2+b^2=\lambda^2$,
where $\lambda$ is a constant.
When $\lambda\ne 0$, we introduce the parametrization
$a=\lambda\cos\theta$
and
$b=\lambda\sin\theta$,
we find that ${c=\frac i2\theta_z}$ and
${\theta_{z\tz}=4k\lambda\sin\theta}$.
Finally, by rescaling $z$ and $\tilde z$, without loss of generality we take $k=\lambda=1/2$.
So the sine-Gordon reduction is
$$
A_z=\frac{i\theta_z}{2}
	\left(
		\begin{matrix}
			1 & 0 \\
			0 & -1
		\end{matrix}
	\right)
,\quad
A_w=\frac 12
	\left(
		\begin{matrix}
			0 & \exp (-i\theta) \\
			\exp(i\theta) & 0
		\end{matrix}
	\right),
$$
$$
\Atz=0
,\quad
\Atw=\frac 12
	\left(
		\begin{matrix}
			0 & 1 \\
			1 & 0
		\end{matrix}
	\right),
$$
where $\theta\equiv\theta(z,\tz)$ solves the sine-Gordon equation
\begin{equation}
\label{sg}
\theta_{z\tz}=\sin\theta.
\end{equation}

In order to construct the B\"acklund transformation, we first construct $J$, and hence $H$ and $K$.
From
$\pw H=- A_wH$ and $\pz H=- A_zH$,
we have
$$
H=
\left(
	\begin{matrix}
		{\rm e}^{i\theta/2} & 0 \\
		0 & {\rm e}^{-i\theta/2} 
	\end{matrix}
\right)
\left(
	\begin{matrix}
		\cosh (w/2) & -\sinh(w/2) \\
		-\sinh (w/2) & \cosh (w/2)
	\end{matrix}
\right)
\widetilde M(\tz,\tw)
$$
and from
$\ptz K=- \Atz K$
and
$\ptw K=- \Atw K$,
we have
$${
K=\exp\left\{-\frac {\tw}2\left(
	\begin{matrix}
		0 & 1 \\
		1 & 0
	\end{matrix}
\right)
\right\}
M(z,w)^{-1}},
$$
where $M(z,w)$ and $\widetilde M(\tz,\tw)$ are in $SL(2;\mathbb{C})$.

Substituting 
$${
J=
F(\tilde w)
\left(
	\begin{matrix}
		{\rm e}^{i\theta/2} & 0 \\
		0 & {\rm e}^{-i\theta/2} 
	\end{matrix}
\right)
F(w)^{-1}
},
$$
where 
\begin{equation}
\label{Fdef}
F(x)=
\exp\left\{\frac {x}2\left(
	\begin{matrix}
		0 & 1 \\
		1 & 0
	\end{matrix}
\right)
\right\}
=
\left(
	\begin{matrix}
		\cosh (x/2) & \sinh (x/2) \\
		\sinh (x/2) & \cosh (x/2)
	\end{matrix}
\right),
\end{equation}
into equations (\ref{BT})
with 
$${
C=\left(
	\begin{matrix}
		a & b \\
		b & a
	\end{matrix}
\right)
}
\quad\mbox{and}\quad
{
\widetilde C=\left(
	\begin{matrix}
		\tilde a & \tilde b \\
		\tilde b & \tilde a
	\end{matrix}
\right)},
$$
gives
$$\begin{array}{ll}
\displaystyle
\tilde a \partial_{\tilde z}(\hat\theta -\theta)= 2 b\sin\left(\frac{\hat\theta+\theta}2\right),&
\displaystyle
\tilde b \partial_{\tilde z}(\hat\theta +\theta)= 2 a\sin\left(\frac{\hat\theta-\theta}2\right),\\
& \\
\displaystyle
a \partial_z(\hat\theta -\theta)= 2\tilde b\sin\left(\frac{\hat\theta+\theta}2\right),&
\displaystyle
b \partial_{z}(\hat\theta +\theta)= 2\tilde a\sin\left(\frac{\hat\theta-\theta}2\right).
\end{array}
$$
Compatibility implies that either $a=\tilde b=0$ or $b=\tilde a=0$. 
These are the standard B\"acklund transformations for the sine-Gordon equation.

\subsection{Reduction to  the sixth Painlev\'e equation}
Each of the six Painlev\'e equations,
\begin{eqnarray}
        u'' &=& 6u^2+z,\\
        u'' &=& 2u^3+zu+\alpha,\\
        u'' &=& \frac 1uu'^2-{1\over z}u'+{1\over z}(\alpha u^2+\beta)+\gamma u^3
+{\delta\over u},\\
        u'' &=& \frac 1{2u}u'^2+\frac 32u^3+4zu^2+2(z^2-\alpha) u+\frac\beta u,
\end{eqnarray}
\begin{eqnarray}
        u'' &=& \left\{\frac 1{2u}+\frac 1{u-1}\right\}u'^2
-\frac 1zu'\nonumber\\
& &+\frac{(u-1)^2}{z^2}\left(\alpha u+\frac\beta u\right)+\frac{\gamma u}z+
{\delta u(u+1)\over u-1},\\
        u'' &=& \frac12\left\{\frac 1u+\frac 1{u-1}+\frac 1{u-z}\right\}u'^2-
\left\{\frac 1z+\frac 1{u-1}+\frac 1{u-z}\right\}u'\nonumber\\
        & &  + {u(u-1)(u-z)\over z^2(z-1)^2}
\left\{\alpha+{\beta z\over u^2}+{\gamma (z-1)\over (u-1)^2}+{\delta z(z-1)\over (u-z)^2}
\right\},\label{pvi}
\end{eqnarray}
where $\alpha$, $\beta$, $\gamma$ and $\delta$ are constants, is known to be a reduction of the ASDYM equations with Lie algebra ${\mathfrak sl}(2;{\mathbb C})$ (Mason and Woodhouse \cite{masonw:93,masonw:96}).  Here we will describe the reduction to the sixth Painlev\'e equation,
${\rm P}_{\rm VI}$ (equation \ref{pvi}).

In terms of the variables
$p=-\log w$, $q=-\log\tilde z$, $r=\log (\tilde w/\tilde z)$, and $t=(z\tilde z)/(w\tilde w)$,
we consider the reduction in which the connection one-form takes the form
$A= P(t){\rm d}p+Q(t){\rm d}q+R(t){\rm d}r$, where $P(t)$, $Q(t)$ and $R(t)$ are functions of $t$ only.  We have
\begin{equation*}
\begin{split}
{\bf A}&=\Az\,{\rm d}z+\Aw\,{\rm d}w+\Atz\,{\rm d}\tz+\Atw\,{\rm d}\tw
\\
&=
P{\rm d}p+Q{\rm d}q+R{\rm d}r
\\
&=
-\frac 1wP{\rm d}w-\frac 1{\tilde z}Q{\rm d}\tilde z+R\left(\frac{{\rm d}\tilde w}{\tilde w}-\frac{{\rm d}\tilde z}{\tilde z}\right).
\end{split}
\end{equation*}
Hence $zA_z=0$, $wA_w=-P$, $\tilde zA_{\tilde z}=-(Q+R)$ and $\tilde wA_{\tilde w}=R$.

The ASDYM equations (\ref{asdym})
reduce to the system of three matrix-valued ODEs
\begin{equation}
\label{matrixpvi}
{P'=0},\quad
{tQ'=[R,Q]}
\mbox{ \  \ and\ \  }
 {t(1-t)R'=[tP+Q,R]},
\end{equation}
where prime denotes differentiation with respect to $t$.  It follows from these equations that the traces of $P^2$, $Q^2$, $R^2$ and $(P+Q+R)^2$ are all constants.  These are related to the constants $\alpha$, $\beta$, $\gamma$ and $\delta$ appearing in ${\rm P}_{\rm VI}$.
Furthermore, on introducing the scaled spectral parameter $\lambda=-\tilde z/(w\zeta)$ and taking $\Psi(z,w,\tilde z,\tilde w;\zeta)=\Phi(t;\lambda)$, we can extend this reduction to the Lax pair (\ref{lax}), giving
\begin{equation*}
\begin{split}
{\partial_t\Phi }&{=-\left(\frac{R}{\lambda-t}\right)\Phi},\\
{\partial_\lambda\Phi }&{=\left(\frac Q\lambda-\frac{P+Q+R}{\lambda-1}+\frac R{\lambda-t}\right)\Phi}.
\end{split}
\end{equation*}

The second equation in (\ref{matrixpvi}) shows that there is an $SL(2;\mathbb{C})$-valued function of $t$, $G(t)$, and a constant $Q_0\in {\mathfrak sl}(2;{\mathbb C})$, such that
$$
{
Q(t)=G(t)^{-1}Q_0G(t)}\  \ \mbox{and}\ \  
{R(t)=-tG(t)^{-1}G'(t)}.
$$ 
The form of $J$ is then
$${
J={\tilde z}^{-Q_0}G(t)w^P}.
$$
From the last equation in (\ref{matrixpvi}), we see that the ASDYM equations in this reduction become
$$
(1-t)(tG^{-1}G')'=[tP+G^{-1}Q_0G,G^{-1}G'].
$$
In the general case, we can take the constant matrices $P$ and $Q_0$ to have the form
$$
P=\frac{\theta_\infty}{2}\left(
		\begin{matrix}
			1 & 0 \\
			0 & -1
		\end{matrix}
	\right)\qquad\mbox{and}\qquad
Q_0=\frac{\theta_0}{2}\left(
		\begin{matrix}
			1 & 0 \\
			0 & -1
		\end{matrix}
	\right	).
$$
So ${J=U(\tilde z)^{-1}G(t)V(w)}$, where ${\displaystyle t=\frac{z\tilde z}{w\tilde w}}$
and
$$
{
U(\tilde z)=\tilde z^{Q_0}=\left(
		\begin{matrix}
			\tilde z^{\theta_0/2} & 0 \\
			0 &\tilde z^{-\theta_0/2}
		\end{matrix}
	\right)},
	\qquad
{
V(w)=w^P\left(
		\begin{matrix}
			w^{\theta_\infty/2} & 0 \\
			0 & w^{-\theta_\infty/2}
		\end{matrix}
	\right)}.
$$
The parameters in ${\rm P}_{\rm VI}$ (equation \ref{pvi}) are given by $\alpha=\frac 12(\theta_\infty-1)^2$, $\beta=-\frac 12\theta_0^2$, $\gamma={\rm tr}\{(P+Q+R)^2\}$
and $\delta={\rm tr}(R^2)+\frac 32$.

Mu\u{g}an and Sakka \cite{mugans:95} derived 
12 Schlesinger transformations for $P_{\,\rm VI}$.   All of these Schlesinger transformations follow from the ASDYM B\"acklund transformation (\ref{BT}) with simple choices of $C$ and $\widetilde C$.  For example, the first Schlesinger transformation in  \cite{mugans:95}, which leaves the parameters $\gamma$ and $\delta$ unchanged and acts on $\alpha$ and $\beta$ via 
$\theta\mapsto\hat \theta_\infty=\theta_\infty+1$ and $\theta\mapsto \hat\theta_0=\theta_0+1$, corresponds to the choice
$$
{
C_1=w^{1/2}\left(
		\begin{matrix}
			0 & 0 \\
			0 & 1
		\end{matrix}
	\right)},
	\quad
	{
	\widetilde C_1=\tilde z^{1/2}\left(
		\begin{matrix}
			0 & 0 \\
			0 & 1
		\end{matrix}
	\right)}.
$$
A self-contained derivation of the Schlesinger transformations for the Painlev\'e equations will be published separately.

\section{Bianchi permutability for the ASDYM equations}


\noindent
In terms of $C$ and $\widetilde C$, the transformation (\ref{darboux}) takes the form
$$
\Psi\mapsto \hat\Psi=(S+\zeta T)\Psi
=
(\hat H\widetilde C H^{-1}+\zeta \hat K C K^{-1})\Psi.
$$
Set $\Phi=K^{-1}\Psi$.  Then, since $J=K^{-1}H$, we have
$$
\Phi\mapsto\hat\Phi=\left(
\hat J\widetilde C J^{-1}+\zeta C
\right)\Phi.
$$
Now suppose that we have two classes of BTs given by the pairs of functions $C^{(1)}$, $\widetilde C^{(1)}$ and $C^{(2)}$, $\widetilde C^{(2)}$.  Now we impose the condition that these two B\"acklund transformations commute.  To this end, let $J_{m,n}$ be a solution of equation (\ref{yang}), with corresponding eigenfunction $\Phi_{m,n}$.
Suppose that under a type 1 B\"acklund transformation, the pair $(J_{m,n},\Phi_{m,n})$ is mapped to $(J_{m+1,n},\Phi_{m+1,n})$ and under a type 2 B\"acklund transformationit is mapped to $(J_{m,n+1},\Phi_{m,n+1})$.
Requiring that this can be done consistently for all integer $m$ and $n$ (i.e., requiring that a type 2 BT applied to $(J_{m+1,n},\Phi_{m+1,n})$ results in the same solution as a type 1 BT applied to  $(J_{m,n+1},\Phi_{m,n+1})$, namely  $(J_{m+1,n+1},\Phi_{m+1,n+1})$)
demands the compatibility of the system
\begin{equation}
\label{auto-lax}
\begin{split}
\Phi_{m+1,n}&=\left(
J_{m+1,n}\widetilde C^{(1)} J_{m,n}^{-1}+\zeta C^{(1)}
\right)\Phi_{m,n},\\
\Phi_{m,n+1}&=\left(
J_{m,n+1}\widetilde C^{(2)} J_{m,n}^{-1}+\zeta C^{(2)}
\right)\Phi_{m,n}.
\end{split}
\end{equation}
Compatibility gives
\begin{equation}
\label{auto}
\begin{split}
&
J^{-1}_{m+1,n+1}\left(
C^{(2)}J_{m+1,n}\widetilde C^{(1)}
-
C^{(1)}J_{m,n+1}\widetilde C^{(2)}
\right)
\\
&
+
\left(
\widetilde C^{(2)}J^{-1}_{m+1,n}C^{(1)}
-
\widetilde C^{(1)} J^{-1}_{m,n+1}C^{(2)}
\right) J_{m,n}
=0,
\end{split}
\end{equation}
together with
$[C^{(1)},C^{(2)}]=[\widetilde C^{(1)},\widetilde C^{(2)}]=0$.
Equations (\ref{auto-lax}) form a Lax pair for the Bianchi permutability equation (\ref{auto}).

Chau and Chinea \cite{chauc:86} previously considered the Bianchi permutatbility for the special B\"acklund transformations derived in Prasad, Sinha and Chau Wang \cite{prasadsw:79}.

\subsection{Sine-Gordon permutability as a reduction}

Substituting
$$
J_{m,n}=
F(\tilde w)
\left(
	\begin{matrix}
		{\rm e}^{i\theta_{m,n}/2} & 0 \\
		0 & {\rm e}^{-i\theta_{m,n}/2} 
	\end{matrix}
\right)
F(w)^{-1},
$$
where $F$ is given by (\ref{Fdef}),
into the permutability equation (\ref{auto})
with
$$
{
C^{(j)}=b_j\left(
		\begin{matrix}
			0 & 1 \\
			1 & 0
		\end{matrix}
	\right)},\ \ 
	{
\widetilde C^{(j)}=\tilde a_j\left(
		\begin{matrix}
			1 & 0 \\
			0 & 1
		\end{matrix}
	\right)}
$$
gives
\begin{equation*}
\begin{split}
&
\sin\left(\frac{\theta_{m+1,n}+\theta_{m,n}}{2}\right)-\sin\left(\frac{\theta_{m,n+1}+\theta_{{m+1,n+1}}}{2}\right)
\\
&
+
\kappa\left[\sin\left(\frac{\theta_{m+1,n}+\theta_{m+1,n+1}}{2}\right)-\sin\left(\frac{\theta_{m,n+1}+\theta_{m,n}}{2}\right)\right]
=0.
\end{split}
\end{equation*}
This is equivalent to the standard form
$$
\tan\left(
\frac
{\theta_{{m+1,n+1}}-\theta_{m,n}}
4
\right)
=
\frac{\kappa+1}{\kappa-1}
\tan\left(
\frac
{\theta_{{m,n+1}}-\theta_{m+1,n}}
4
\right),
$$
of the Bianchi permutability theorem for the sine-Gordon equation.

\subsection{Non-autonomous ASDYM Bianchi system}

A non-autonomous version of the Bianchi system can be derived by allowing the matrices $C^{(j)}$ and $\widetilde C^{(j)}$, $j=1,2$, to depend on $m$ and $n$.  To this end, we replace the system (\ref{auto-lax}) with the system
\begin{equation*}
\begin{split}
\Phi_{m+1,n}&=\left(
J_{m+1,n}\widetilde C^{(1)}_{m,n} J_{m,n}^{-1}+\zeta C^{(1)}_{m,n}
\right)\Phi_{m,n},\\
\Phi_{m,n+1}&=\left(
J_{m,n+1}\widetilde C^{(2)}_{m,n} J_{m,n}^{-1}+\zeta C^{(2)}_{m,n}
\right)\Phi_{m,n}.
\end{split}
\end{equation*}
Compatibility gives
\begin{equation}
\label{nonauto}
\begin{split}
&
J^{-1}_{m+1,n+1}\left(
C^{(2)}_{m+1,n}J_{m+1,n}\widetilde C^{(1)}_{m,n}
-
C^{(1)}_{m,n+1}J_{m,n+1}\widetilde C_{m,n}^{(2)}
\right)
\\
&
+
\left(
\widetilde C^{(2)}_{m+1,n} J^{-1}_{m+1,n}C^{(1)}_{m,n}
-
\widetilde C^{(1)}_{m,n+1} J^{-1}_{m,n+1}C^{(2)}_{m,n}
\right) J_{m,n}
=0,
\end{split}
\end{equation}
where
\begin{equation}
\label{Ccond}
C^{(1)}_{m,n+1}C^{(2)}_{m,n}=C^{(2)}_{m+1,n}C^{(1)}_{m,n}
\end{equation}
and
\begin{equation}
\label{tCcond}
\widetilde C^{(1)}_{m,n+1}\widetilde C^{(2)}_{m,n}=\widetilde C^{(2)}_{m+1,n}\widetilde C^{(1)}_{m,n}.
\end{equation}
This is a nonautonomous version of the Bianchi system (\ref{auto}).

\subsection{Reduction of Bianchi system}
Let us consider the system 
(\ref{nonauto}--\ref{tCcond})
independently of any connection with the ASDYM equations.
Let
$$
C^{(1)}_{m,n}=\frac1{\alpha_m}\left(
	\begin{matrix}
		1 & 0\\
		0 & 1
	\end{matrix}
\right),\quad
C^{(2)}_{m,n}=\frac1{\beta_n}\left(
	\begin{matrix}
		1 & 0\\
		0 & 1
	\end{matrix}
\right),
\quad
\widetilde C^{(1)}_{m,n}=\widetilde C^{(2)}_{m,n}=
\left(
	\begin{matrix}
		0 & 1\\
		1 & 0
	\end{matrix}
\right),
$$
and
$$
J_{m,n}=\left(
	\begin{matrix}
		u_{m,n} & 0\\
		0 & 1/u_{m,n}
	\end{matrix}
\right).
$$
Then the nonautonomous Bianchi system (\ref{nonauto}) reduces to
\begin{equation}
\label{lmkdv}
\begin{split}
&
\alpha_m(u_{m,n}u_{m+1,n}-u_{m,n+1}u_{m+1,n+1})
\\
&
-\beta_n(u_{m,n}u_{m,n+1}-u_{m+1,n}u_{m+1,n+1})=0.
\end{split}
\end{equation}
Equation (\ref{lmkdv}) is known as the nonautonomous lattice mKdV equation \cite{papageorgiougr:93}.
In order to try to interpret this in terms of B\"acklund transformations, note that symmetry reductions of the ASDYM equations in the original variables ($A_\mu$) lead to reductions in Yang's form where $J$ has the ``dressed'' form ${J=AGB}$, where $G$ depends on the variables that will be the independent variables of the reduced equations and the ``dressing matrices'' $A$ and $B$ depend on some auxiliary combination of variables that do not appear in the reduced equation.

Substituting the form ${J_{m,n}=AG_{m,n}B}$ into the nonautonomous Bianchi permutability equation (\ref{nonauto}) gives
$$
{
G^{-1}_{m+1,n+1}\left(
D^{(2)}_{m+1,n}G_{m+1,n}\widetilde D^{(1)}_{m,n}
-
D^{(1)}_{m,n+1}G_{m,n+1}\widetilde D_{m,n}^{(2)}
\right)
}
$$
$$
{
+
\left(
\widetilde D^{(2)}_{m+1,n} G^{-1}_{m+1,n}D^{(1)}_{m,n}
-
\widetilde D^{(1)}_{m,n+1} G^{-1}_{m,n+1}D^{(2)}_{m,n}
\right) G_{m,n}
=0
},
$$
where
${D^{(j)}=A^{-1}C^{(j)}A}$
and
${\widetilde D^{(j)}=B \widetilde C^{(j)}B^{-1}}$.

Furthermore, the conditions
$$
{
C^{(1)}_{m,n+1}C^{(2)}_{m,n}=C^{(2)}_{m+1,n}C^{(1)}_{m,n}
}
\mbox{\ \ \ and\ \ \ }
{
\widetilde C^{(1)}_{m,n+1}\widetilde C^{(2)}_{m,n}=\widetilde C^{(2)}_{m+1,n}\widetilde C^{(1)}_{m,n}
}
$$
become
$$
{
D^{(1)}_{m,n+1}D^{(2)}_{m,n}=D^{(2)}_{m+1,n}D^{(1)}_{m,n}
}
\mbox{\ \ \ and\ \ \ }
{
\widetilde D^{(1)}_{m,n+1}\widetilde D^{(2)}_{m,n}=\widetilde D^{(2)}_{m+1,n}\widetilde D^{(1)}_{m,n}
}.
$$

Now we return to the problem of interpreting the nonautonomous lattice mKdV (\ref{lmkdv}) in terms of B\"acklund transformations.
The ASDYM Bianchi system (\ref{nonauto}) with
$${
J_{m,n}=
F(\tilde w)
\left(
	\begin{matrix}
		u_{m,n}(z,\tilde z) & 0 \\
		0 & \frac 1{u_{m,n}(z,\tilde z)}
	\end{matrix}
\right)
F(w)^{-1}
}
$$
and
$$
C^{(1)}_{m,n}=\frac1{\alpha_m}\left(
	\begin{matrix}
		1 & 0\\
		0 & 1
	\end{matrix}
\right),\quad
C^{(2)}_{m,n}=\frac1{\beta_n}\left(
	\begin{matrix}
		1 & 0\\
		0 & 1
	\end{matrix}
\right),
\quad
\widetilde C^{(1)}_{m,n}=\widetilde C^{(2)}_{m,n}=
\left(
	\begin{matrix}
		0 & 1\\
		1 & 0
	\end{matrix}
\right),
$$
again gives the nonautonomous lattice mKdV (\ref{lmkdv}).
However this is now a statement about BTs of a reduction of the ASDYM equations (specifically the sine-Gordon equation with $u_{m,n}(z,\tilde z)={\rm e}^{i\theta_{m,n}(z,\tilde z)/2}$).
Ormerod \cite{ormerod:12} has shown that dmKdV has a reduction to qP$_{\rm{VI}}$.

\section{Conclusion}
Many integrable equations are known to be reductions of the ASDYM equations.
We have derived a class of B\"acklund transformations (\ref{BT}) for these equations that depends on two functions $C(z,w)$ and $\widetilde C(\tilde z,\tilde w)$.
This form is particularly useful when we consider reductions.  In particular the Schlesinger transformations for the Painlev\'e equations follow from the standard reductions from the ASDYM equations together, where symmetry considerations restrict the possible choices for $C$ and $\widetilde C$.  

The Bianchi permutability of the B\"acklund transformations (\ref{BT}) results in the system (\ref{nonauto}).
This is a discrete integrable system in its own right.  Through various reductions, this is a rich source of discrete systems.  It also provides a way of identifying certain discrete equations as B\"acklund transformations of integrable differential systems.
The richness of reductions of the ASDYM equations comes form the large 
(conformal) group of symmetries.
The richness of discrete reductions of the ASDYM Bianchi system comes from the choices of $C(z,w)$, $\widetilde C(\tilde z,\tilde w)$ as well as the form of $J$.

\section*{Acknowledgments}
RH gratefully acknowledges support from the Engineering and Physical Science Research Council through grant EP/K041266/1.


\end{document}